\documentclass{article}

\PassOptionsToPackage{numbers, compress}{natbib}
 \usepackage[dblblindworkshop, final]{neurips_2025}
\workshoptitle{Machine Learning and the Physical Sciences}

\usepackage[utf8]{inputenc} % allow utf-8 input
\usepackage[T1]{fontenc}    % use 8-bit T1 fonts
\usepackage{hyperref}       % hyperlinks
\usepackage{url}            % simple URL typesetting
\usepackage{booktabs}       % professional-quality tables
\usepackage{amsfonts}       % blackboard math symbols
\usepackage{nicefrac}       % compact symbols for 1/2, etc.
\usepackage{microtype}      % microtypography
\usepackage{xcolor}         % colors
\usepackage{cite}
\usepackage{graphicx}
\usepackage[export]{adjustbox}
\title{Latent Representation Learning in Heavy-Ion Collisions with MaskPoint Transformer}

\author{%
Jing-Zong Zhang and Shuang Guo\\
Institute of Modern Physics, Fudan University\\
Shanghai 200433, China\\
Shanghai Research Center for Theoretical Nuclear Physics\\
NSFC and Fudan University, Shanghai 200438, China\\
\And
Li-Lin Zhu\\
Department of Physics, Sichuan University\\
Chengdu 610064, China\\
\And
Lingxiao Wang\\
RIKEN Interdisciplinary Theoretical and Mathematical Sciences (iTHEMS), \\ Wako, Saitama 351-0198, Japan \\
\texttt{lingxiao.wang@riken.jp}\\
\And
Guo-Liang Ma\\
Institute of Modern Physics, Fudan University\\
Shanghai 200433, China\\
Shanghai Research Center for Theoretical Nuclear Physics\\
NSFC and Fudan University, Shanghai 200438, China\\
\texttt{glma@fudan.edu.cn}
}

\begin{document}

\maketitle

\begin{abstract}

A central challenge in high-energy nuclear physics is to extract informative features from the high-dimensional final-state data of heavy-ion collisions (HIC) in order to enable reliable downstream analyses. Traditional approaches often rely on selected observables, which may miss subtle but physically relevant structures in the data. To address this, we introduce a Transformer-based autoencoder trained with a two-stage paradigm: self-supervised pre-training followed by supervised fine-tuning. The pretrained encoder learns latent representations directly from unlabeled HIC data, providing a compact and information-rich feature space that can be adapted to diverse physics tasks. As a case study, we apply the method to distinguish between large and small collision systems, where it achieves significantly higher classification accuracy than PointNet. Principal component analysis and SHAP interpretation further demonstrate that the autoencoder captures complex non-linear correlations beyond individual observables, yielding features with strong discriminative and explanatory power. These results establish our two-stage framework as a general and robust foundation for feature learning in HIC, opening the door to more powerful analyses of quark–gluon plasma properties and other emergent phenomena. The implementation is publicly available at \url{https://github.com/Giovanni-Sforza/MaskPoint-AMPT}.
\end{abstract}

\section{Introduction}

Relativistic heavy-ion collisions provide a unique environment to study the QCD phase transition of nuclear matter at extreme temperature and density, and to probe the properties of deconfined partonic matter, known as quark–gluon plasma (QGP) ~\citep{zong1,zong2,ALICE:2008ngc,Bzdak:2019pkr,Luo:2017faz}. Traditional observables such as particle spectra and anisotropic flows have revealed essential QGP features ~\citep{zong10,zong11,zong12,zong13}, yet they are limited in dealing with high-dimensional final-state data. This restricts their effectiveness in capturing the subtle dynamical differences across varying collision system sizes, for example, between large (Pb+Pb) and small (p+Pb) systems~\citep{Dusling:2015gta,Loizides:2016tew,Nagle:2018nvi,Wang:2022rdh,Zhao:2021bef,Pei:2025jey,Ma:2016bbw,Bzdak:2014dia}.

Recent advances in machine learning have introduced deep learning techniques into high-energy nuclear physics ~\citep{Graczykowski:2022zae,Meng:2022ssp,LingXiao:2022ssp,WanBing:2022ssp,YiLun:2022ssp,fupeng:2022ssp,He:2023urp,Ma:2023zfj,Zhou:2023pti,Boehnlein:2021eym,Aarts:2025gyp}. Since the final-state particle phase space can be naturally represented as unordered point clouds, models such as PointNet and Point Transformer ~\citep{point-m2ae,2d23d,point-bert,maskpoint} have been applied to extract representations directly from their raw data~\citep{Zhou:2023pti}. These approaches enable data-driven exploration of QGP properties beyond traditional observables~\citep{golling2024masked, birk2024omnijet, harris2024re, mikuni2025method}.

In this work, we introduce a Transformer-based autoencoder ~\citep{maskpoint} into heavy-ion collision analysis, employing a two-stage paradigm of self-supervised pretraining followed by supervised finetuning. The training data are provided by the AMPT (A Multi-Phase Transport) model ~\citep{lin2005multiphase,lin2021further}, a widely used framework that incorporates initial condition, parton cascade, hadronization, and hadronic rescatterings, and has been shown to reproduce many experimental observables. Using Pb+Pb and p+Pb events generated by AMPT, the autoencoder first learns latent representations by minimizing reconstruction error. The pretrained encoder is then finetuned with a lightweight classifier for system identification. Compared with the PointNet baseline \citep{pointnet,guo2024machine}, our method achieves a significantly higher classification accuracy. Furthermore, principal component analysis, correlation analysis, and SHAP interpretation reveal that the learned features capture both linear and nonlinear connections with traditional observables. {While identifying the collision system is not a practical challenge, we employ it as a benchmark task to quantitatively assess the discriminative power of the latent representations learned by our model\citep{lai2022information}.}

Our results show that self-supervised pretraining with a Transformer-based autoencoder can extract physically meaningful structures from high-dimensional collision data. This two-stage framework offers an effective methodology for future studies of QGP properties.

\section{Model Architecture and Methodology}

\begin{figure}[htbp]
  \centering
  %\fbox{\rule[-.5cm]{0cm}{4cm} \rule[-.5cm]{4cm}{0cm}}
  \includegraphics[width=0.75\textwidth]{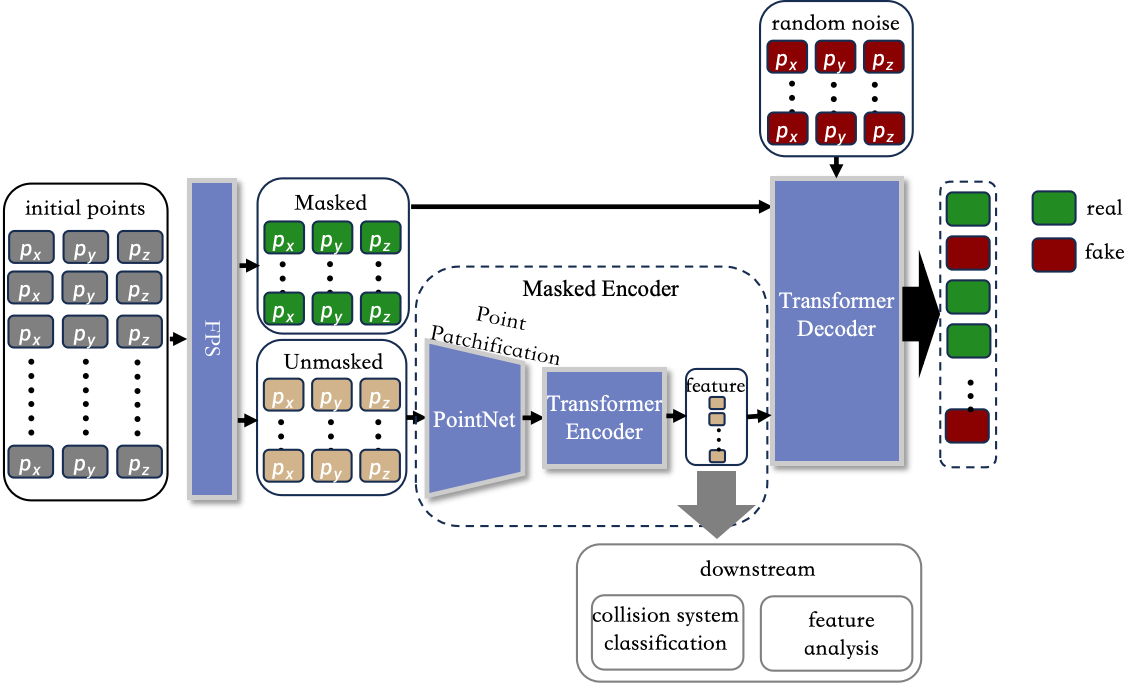}
  \caption{Architecture of the masked autoencoder. The input point cloud is first partitioned into patches by PointNet and then encoded by a Transformer. During pre-training, the encoder output is passed to a Transformer decoder for an discrimination task between real and fake particles. For fine-tuning and analysis, only the pretrained encoder is retained as the feature extractor.}\label{fig.autoencoder}
\end{figure}

This study utilizes p+Pb and Pb+Pb collision events from the AMPT model. Each event is represented as a point cloud composed of the three-momenta ($p_{\rm x}, p_{\rm y}, p_z$) of 128 final-state particles within the kinematic window of pseudorapidity $|\eta|<2.4$ and transvese momentum $p_{\rm T} > 0.4$ GeV/\textit{c} \citep{CMS:2013jlh,guo2024machine}. To learn the intrinsic physical structure of the data, rather than degenerating into a trivial coordinate-memorization task, we designed a masked autoencoder based on a discrimination scheme for pre-training (see Figs. \ref{fig.autoencoder}) \citep{maskpoint}. Specifically, we randomly mask 25\% of each event's point cloud using Farthest Point Sampling (FPS), and the unmasked portion, $\mathbf{U} \in \mathbb{R}^{96\times3}$, is fed into an encoder. This encoder first extracts local features using a PointNet network, then captures global information through a 6-layer Transformer architecture, generating a 96-dimensional feature vector $\mathbf{f}$. Subsequently, a single-layer Transformer decoder employs a cross-attention mechanism to integrate the feature vector $\mathbf{f}$ with either the masked "real point cloud" or a randomly sampled "fake point cloud," which is then judged by an MLP binary classifier head {using cross-entropy loss function}. This discrimination task compels the encoder to learn high-quality physical features. After pre-training, we freeze the encoder parameters for downstream tasks. 

For the task of collision system identification, we remove the decoder and feed the features $\mathbf{f}$ extracted by the encoder into a simple MLP classifier. {Both the pre-training and fine-tuning stages were run for up to 300 epochs, with the best-performing model saved based on validation performance.} To analyze the physical meaning of the features, we use Principal Component Analysis for dimensionality reduction, then calculate the linear correlations between the principal components and physical observables. Finally, we combine a Random Forest model with SHAP (SHapley Additive exPlanations) analysis to investigate non-linear associations between them \citep{NIPS2017_8a20a862}.

\section{Results and discussion}

\begin{figure}[htbp]
  \centering
  %\fbox{\rule[-.5cm]{0cm}{4cm} \rule[-.5cm]{4cm}{0cm}}
  \includegraphics[width=0.58\textwidth]{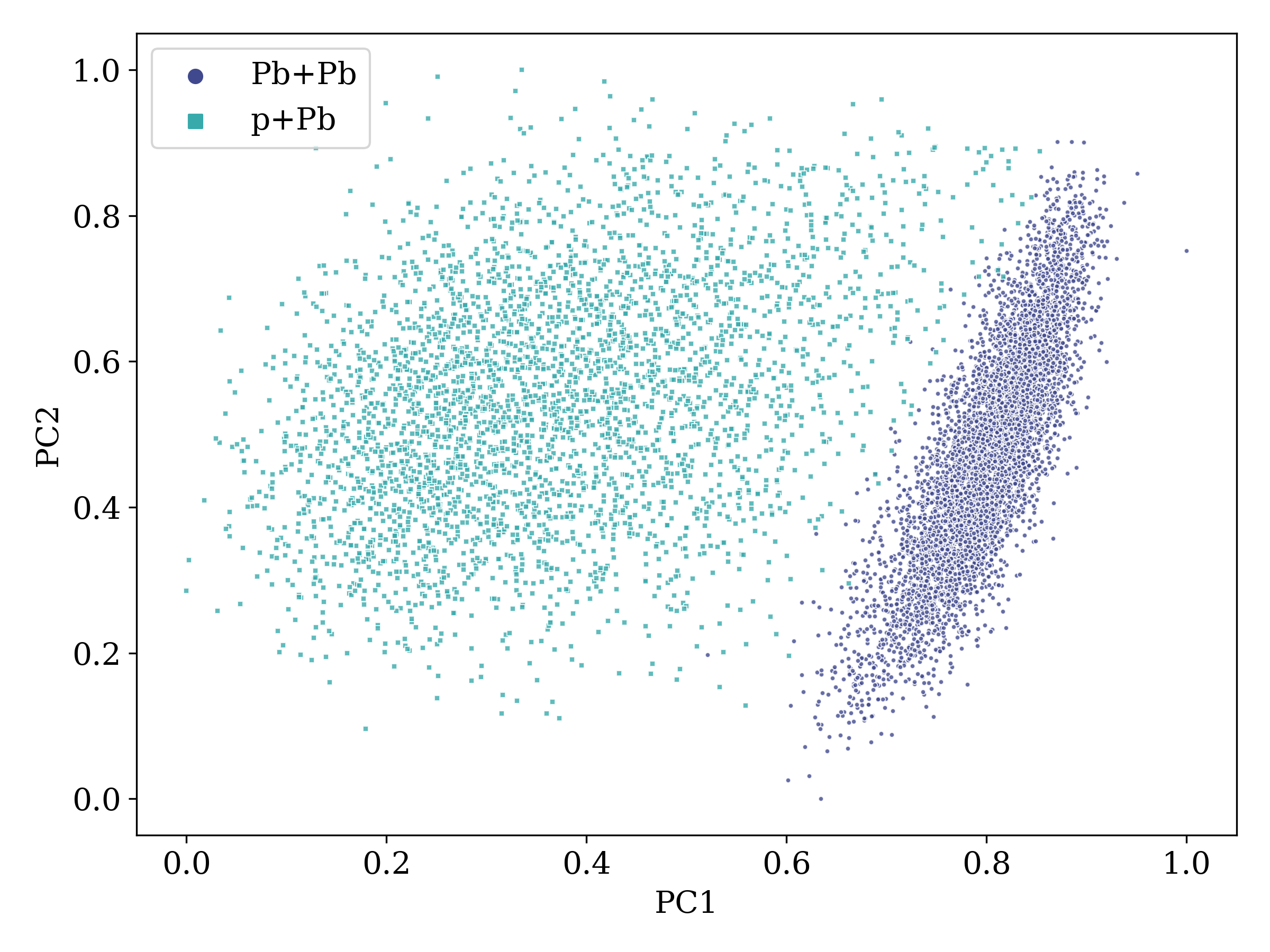}
  \caption{PCA projection of the latent features learned by the autoencoder during pre-training with 3D momentum inputs, shown in the PC1–PC2 plane. The two colors denote two different systems. The clear clustering indicates that the model, even without labels, captures and distinguishes intrinsic physical differences between the two systems.}\label{fig.pretraining_scatter}
\end{figure}

This study demonstrates the successful application of a masked autoencoder within a self-supervised pre-training and supervised fine-tuning paradigm to extract physically meaningful representations from the final-state three-dimensional momentum data of heavy-ion collisions. In the pre-training phase, the model, without any label information, learned to spontaneously distinguish between p+Pb and Pb+Pb collision systems in its latent space (Fig.~\ref{fig.pretraining_scatter}), highlighting the potential of self-supervised learning to uncover intrinsic data structures.

\begin{figure}[htbp]
    \centering
    \begin{minipage}[t]{0.48\textwidth} 
        \centering
        \includegraphics[width=\textwidth]{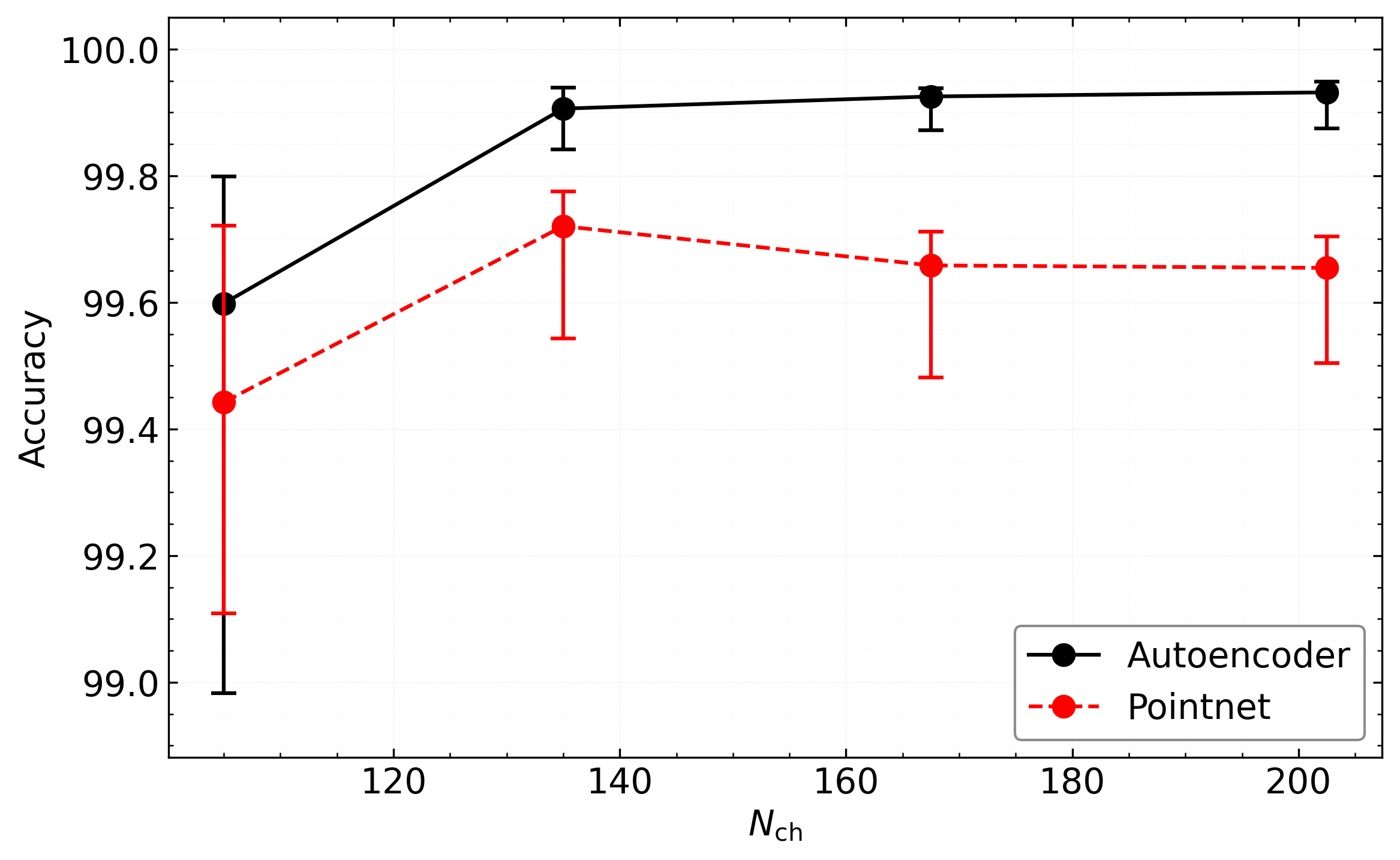}
        \caption{Classification accuracy between large and small systems across $N_{ch}$ bins for the fine-tuned autoencoder (black) and PointNet (red). The autoencoder consistently outperforms PointNet, validating the effectiveness of the “self-supervised pre-training + supervised fine-tuning” strategy.}\label{fig.acc_finture}
    \end{minipage}
    \hfill 
    \begin{minipage}[t]{0.48\textwidth} 
        \centering
        \includegraphics[width=\textwidth]{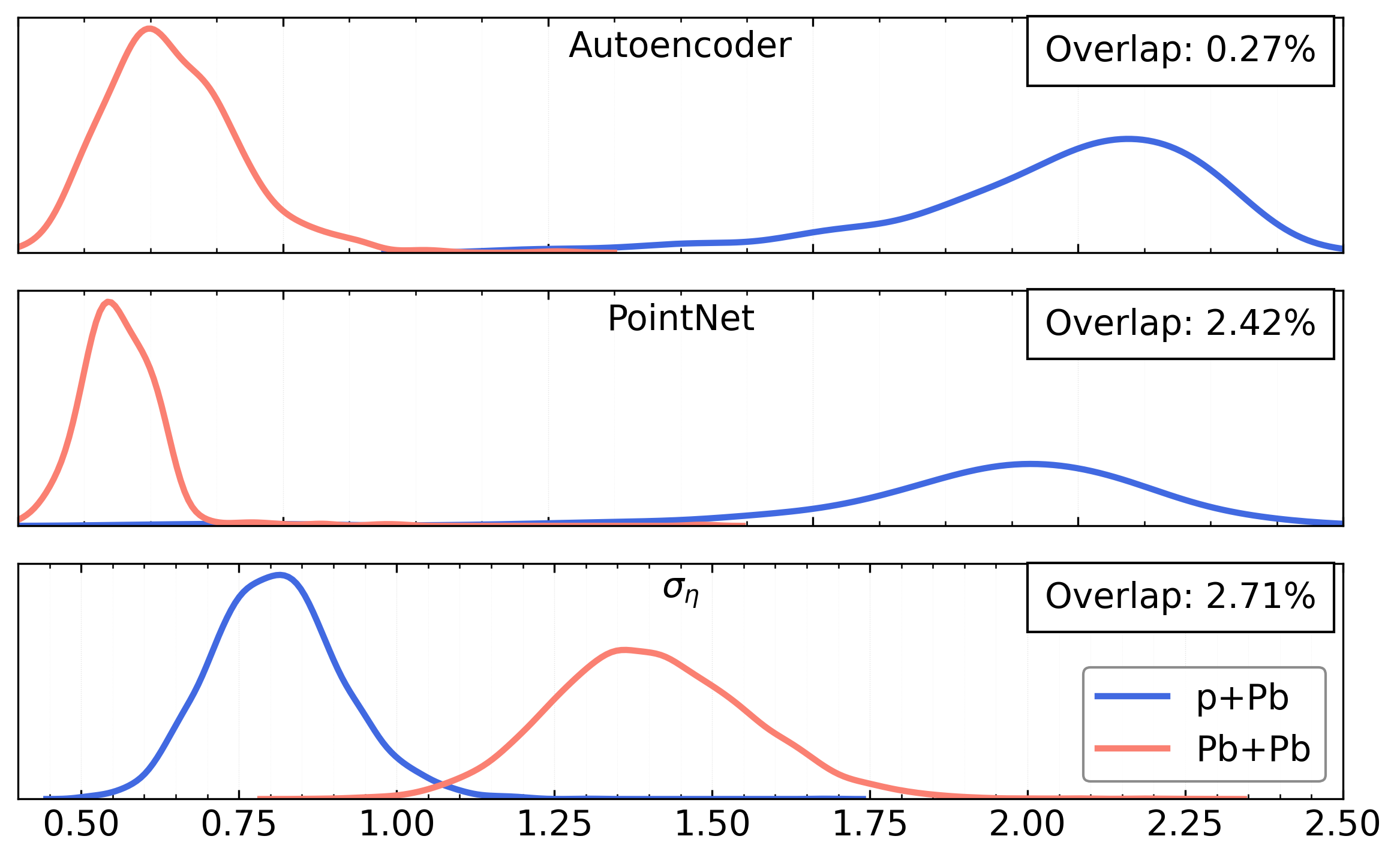}
        \caption{ %Distributions of PC1 from the autoencoder (top) and PointNet (middle), compared with $\sigma_{\eta}$ (bottom). The overlap between Pb+Pb and p+Pb is only 0.27\% for the autoencoder, far below both PointNet (2.42\%) and the physical limit from $\sigma_{\eta}$ (2.71\%). While PointNet merely approaches the theoretical discriminative bound set by $\sigma_{\eta}$, the autoencoder surpasses it, demonstrating that the features learned through self-supervised pre-training provide substantially stronger discriminative power.
         Distributions of PC1 from the autoencoder (top) and PointNet (middle), compared with $\sigma_{\eta}$ (bottom). The overlap between Pb+Pb and p+Pb is 0.27\% for the autoencoder, lower than PointNet (2.42\%) and $\sigma_{\eta}$ (2.71\%). While PointNet approaches the theoretical limit of $\sigma_{\eta}$, the autoencoder surpasses it, demonstrating stronger discriminative power from self-supervised pre-training.}\label{fig.overlap}
    \end{minipage}
\end{figure}

In the subsequent fine-tuning phase, our model's superiority is quantitatively demonstrated by its classification accuracy, which significantly surpasses the PointNet baseline across all tested multiplicity ranges (Fig.~\ref{fig.acc_finture}). Error bars represent the range of the accuracies over 10 independent trials, where the test set is split into 10 subsets for each trial. To investigate the origin of this performance advantage, we analyzed the principal components (PCs) of the features learned by both models. This analysis reveals a striking difference in the discriminative power of their leading PCs (Fig.~\ref{fig.overlap}). The distribution overlap for our autoencoder's PC1 is merely 0.27\%, substantially lower than the 2.42\% overlap for PointNet's PC1. Crucially, PointNet's overlap closely mirrors the 2.71\% overlap of the raw physical observable, the standard deviation of pseudorapidity distribution ($\sigma_{\eta}$), suggesting that its learned representation is closely tied to this single variable.

\begin{figure}[htbp]
  \centering
  %\fbox{\rule[-.5cm]{0cm}{4cm} \rule[-.5cm]{4cm}{0cm}}
  \includegraphics[width=0.9\textwidth]{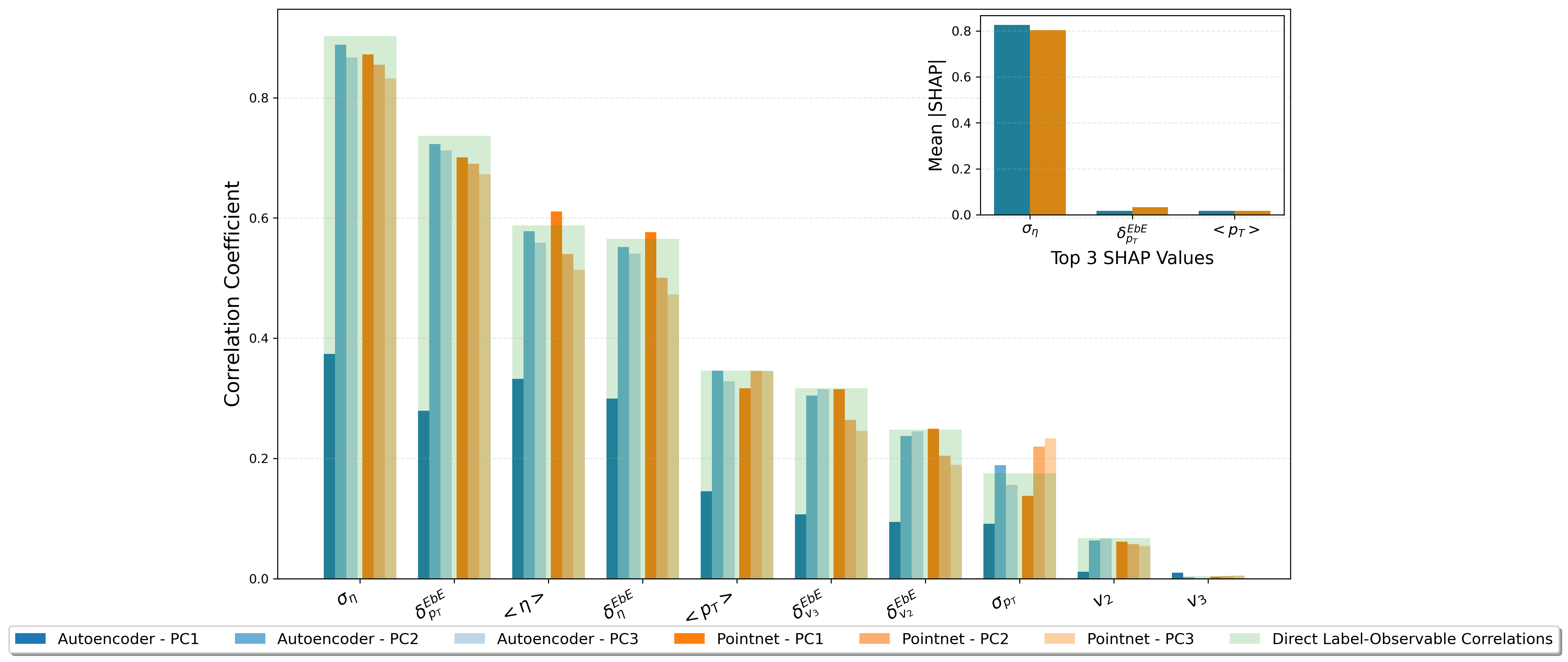}
  \caption{Correlation coefficients between PCs and observables, with SHAP contributions to PC1. Autoencoder PC1 shows near-zero linear correlations but a high SHAP weight from $\sigma_{\eta}$, indicating that it encodes key information in a non-linear manner, which explains its superior performance.}\label{fig.correlation}
\end{figure}

This observation motivates a deeper analysis using SHAP values and correlation coefficients to probe the structure of the learned features. Fig.~\ref{fig.correlation} shows that PointNet’s PC1 is strongly linearly correlated with $\sigma_{\eta}$, indicating that it mainly reproduces a linear representation of the most discriminative observable. By contrast, autoencoder’s PC1 exhibits almost zero linear correlation with $\sigma_{\eta}$, yet SHAP analysis identifies $\sigma_{\eta}$ as its most influential contributor. This combination of low linear correlation and high SHAP importance demonstrates that the autoencoder, through self-supervised pre-training, learns a more complex non-linear representation. Rather than duplicating a single observable, the pretrained encoder extracts richer and more abstract features, which explains its markedly lower distribution overlap and the resulting superior classification performance. Therefore, this work not only provides a high-performance model for identifying heavy ion collision systems, but clearly demonstrates that selfsupervised pretraining can guide a model to move beyond simple linear fitting of key physical observables and instead learn deeper, non-linear patterns within the data, offering a powerful new perspective for AI-driven discovery in fundamental high-energy nuclear physics.

\section{Conclusion, Limitations and Outlook}

This work demonstrates the effectiveness of a self-supervised pre-training and fine-tuning paradigm with a Transformer-based autoencoder for heavy-ion collision analysis. Our model learns deep and physically meaningful representations from unlabeled data, outperforming PointNet by capturing complex non-linear features rather than merely reproducing obvious tranditional  observables.

Despite these advantages, several limitations remain. Our study relies solely on AMPT-generated events and only considers p+Pb and Pb+Pb systems. The input is limited to particle three-momenta $(p_x,p_y,p_z)$, without incorporating four-momentum or additional features such as charge or spin. Addressing these aspects is crucial for enhancing robustness and broad applicability.

Looking forward, promising directions include exploring advanced generative and Transformer variants~\citep{dohi2020variational,ostdiek2022deep,mikuni2023fast,butter2025jet,qu2022particle,qiu2023holistic}, embedding physics-inspired priors such as conservation laws and symmetries~\citep{bogatskiy2022symmetry}, and strengthening the connection between latent representations and established physical observables~\citep{gambhir2024streamlining,hallin2025universal}. {Beyond the generic masking strategy adopted here, pre-training can further be designed with physics-informed objectives to construct data-driven observables~\citep{gong2022efficient,bogatskiy2022pelican,spinner2024lorentz,brehmer2024lorentz,spinner2025lorentz,favaro2025lorentz}}, potentially uncovering subtle signals such as the Chiral Magnetic Effect~\citep{Fukushima:2008xe,Zhao:2021yjo,Guo:2025wry} or nuclear deformation~\citep{Jia:2022qgl,Zhao:2022grq}. Such strategies could transform machine learning into an active partner in the AI discovery of new physics in high-energy heavy-ion collisions.

\section{Acknowledgments}
This work is partially supported by the National Natural Science Foundation of China under Grants No.12147101 and No. 12325507, the National Key Research and Development Program of China under Grant No. 2022YFA1604900, the Guangdong Major Project of Basic and Applied Basic Research under Grant No. 2020B0301030008 (J.Z, S.G. and G.M.), the RIKEN TRIP initiative (RIKEN Quantum), JSPS KAKENHI Grant No. 25H01560, and JST-BOOST Grant No.JPMJBY24H9 (L.W.). We also thank the DEEP-IN working group at RIKEN-iTHEMS for support in the preparation.

\bibliographystyle{jhep}
\bibliography{Manualbib.bib}

\providecommand{\href}[2]{#2}\begingroup\raggedright\begin{thebibliography}{10}

\bibitem{zong1}
J.~Adams, M.~Aggarwal, Z.~Ahammed, J.~Amonett, B.~Anderson, D.~Arkhipkin et~al., \emph{Experimental and theoretical challenges in the search for the quark--gluon plasma: The star collaboration's critical assessment of the evidence from rhic collisions}, {\emph{Nuclear Physics A} {\bfseries 757} (2005) 102}.

\bibitem{zong2}
K.~Adcox, S.S.~Adler, S.~Afanasiev, C.~Aidala, N.~Ajitanand, Y.~Akiba et~al., \emph{Formation of dense partonic matter in relativistic nucleus--nucleus collisions at rhic: experimental evaluation by the phenix collaboration}, {\emph{Nuclear Physics A} {\bfseries 757} (2005) 184}.

\bibitem{ALICE:2008ngc}
{\scshape ALICE} collaboration, \emph{{The ALICE experiment at the CERN LHC}}, \href{https://doi.org/10.1088/1748-0221/3/08/S08002}{\emph{JINST} {\bfseries 3} (2008) S08002}.

\bibitem{Bzdak:2019pkr}
A.~Bzdak, S.~Esumi, V.~Koch, J.~Liao, M.~Stephanov and N.~Xu, \emph{{Mapping the Phases of Quantum Chromodynamics with Beam Energy Scan}}, \href{https://doi.org/10.1016/j.physrep.2020.01.005}{\emph{Phys. Rept.} {\bfseries 853} (2020) 1} [\href{https://arxiv.org/abs/1906.00936}{{\ttfamily 1906.00936}}].

\bibitem{Luo:2017faz}
X.~Luo and N.~Xu, \emph{{Search for the QCD Critical Point with Fluctuations of Conserved Quantities in Relativistic Heavy-Ion Collisions at RHIC : An Overview}}, \href{https://doi.org/10.1007/s41365-017-0257-0}{\emph{Nucl. Sci. Tech.} {\bfseries 28} (2017) 112} [\href{https://arxiv.org/abs/1701.02105}{{\ttfamily 1701.02105}}].

\bibitem{zong10}
F.~Karsch, A.~Bazavov, H.-T.~Ding, P.~Hegde, O.~Kaczmarek, E.~Laermann et~al., \emph{Conserved charge fluctuations from lattice qcd and the beam energy scan}, {\emph{Nuclear Physics A} {\bfseries 956} (2016) 352}.

\bibitem{zong11}
H.-j.~Xu, Z.~Li and H.~Song, \emph{High-order flow harmonics of identified hadrons in 2.76 a tev pb+ pb collisions}, {\emph{Physical Review C} {\bfseries 93} (2016) 064905}.

\bibitem{zong12}
X.~Zhu, Y.~Zhou, H.~Xu and H.~Song, \emph{Correlations of flow harmonics in 2.76 a tev pb-pb collisions}, {\emph{Physical Review C} {\bfseries 95} (2017) 044902}.

\bibitem{zong13}
N.-B.~Chang, S.~Cao and G.-Y.~Qin, \emph{Probing medium-induced jet splitting and energy loss in heavy-ion collisions}, {\emph{Physics Letters B} {\bfseries 781} (2018) 423}.

\bibitem{Dusling:2015gta}
K.~Dusling, W.~Li and B.~Schenke, \emph{{Novel collective phenomena in high-energy proton\textendash{}proton and proton\textendash{}nucleus collisions}}, \href{https://doi.org/10.1142/S0218301316300022}{\emph{Int. J. Mod. Phys. E} {\bfseries 25} (2016) 1630002} [\href{https://arxiv.org/abs/1509.07939}{{\ttfamily 1509.07939}}].

\bibitem{Loizides:2016tew}
C.~Loizides, \emph{{Experimental overview on small collision systems at the LHC}}, \href{https://doi.org/10.1016/j.nuclphysa.2016.04.022}{\emph{Nucl. Phys. A} {\bfseries 956} (2016) 200} [\href{https://arxiv.org/abs/1602.09138}{{\ttfamily 1602.09138}}].

\bibitem{Nagle:2018nvi}
J.L.~Nagle and W.A.~Zajc, \emph{{Small System Collectivity in Relativistic Hadronic and Nuclear Collisions}}, \href{https://doi.org/10.1146/annurev-nucl-101916-123209}{\emph{Ann. Rev. Nucl. Part. Sci.} {\bfseries 68} (2018) 211} [\href{https://arxiv.org/abs/1801.03477}{{\ttfamily 1801.03477}}].

\bibitem{Wang:2022rdh}
H.-S.~Wang and G.-L.~Ma, \emph{{Testing the collectivity in large and small colliding systems with test particles}}, \href{https://doi.org/10.1103/PhysRevC.106.064907}{\emph{Phys. Rev. C} {\bfseries 106} (2022) 064907} [\href{https://arxiv.org/abs/2208.06854}{{\ttfamily 2208.06854}}].

\bibitem{Zhao:2021bef}
X.-L.~Zhao, Z.-W.~Lin, L.~Zheng and G.-L.~Ma, \emph{{A transport model study of multiparticle cumulants in p+p collisions at 13 TeV}}, \href{https://doi.org/10.1016/j.physletb.2023.137799}{\emph{Phys. Lett. B} {\bfseries 839} (2023) 137799} [\href{https://arxiv.org/abs/2112.01232}{{\ttfamily 2112.01232}}].

\bibitem{Pei:2025jey}
J.-L.~Pei, G.-L.~Ma and A.~Bzdak, \emph{{Effect of transverse momentum conservation and flow on symmetric cumulants $sc_{2,3} \left \{ 4 \right \}$ and $sc_{2,3,4} \left \{ 6 \right \}$}},  \href{https://arxiv.org/abs/2503.12846}{{\ttfamily 2503.12846}}.

\bibitem{Ma:2016bbw}
G.-L.~Ma and A.~Bzdak, \emph{{Flow in small systems from parton scatterings}}, \href{https://doi.org/10.1016/j.nuclphysa.2016.01.057}{\emph{Nucl. Phys. A} {\bfseries 956} (2016) 745}.

\bibitem{Bzdak:2014dia}
A.~Bzdak and G.-L.~Ma, \emph{{Elliptic and triangular flow in $p$+Pb and peripheral Pb+Pb collisions from parton scatterings}}, \href{https://doi.org/10.1103/PhysRevLett.113.252301}{\emph{Phys. Rev. Lett.} {\bfseries 113} (2014) 252301} [\href{https://arxiv.org/abs/1406.2804}{{\ttfamily 1406.2804}}].

\bibitem{Graczykowski:2022zae}
{\scshape ALICE} collaboration, \emph{{Using machine learning for particle identification in ALICE}}, \href{https://doi.org/10.1088/1748-0221/17/07/C07016}{\emph{JINST} {\bfseries 17} (2022) C07016} [\href{https://arxiv.org/abs/2204.06900}{{\ttfamily 2204.06900}}].

\bibitem{Meng:2022ssp}
M.~ZHOU, Y.~LUO and H.~SONG, \emph{{Applications of machine learning in relativistic heavy ion physics}}, \href{https://doi.org/https://doi.org/10.1360/SSPMA-2021-0321}{\emph{Sci. China Phys. Mech. Astron.} {\bfseries 52} (2022) 252002}.

\bibitem{LingXiao:2022ssp}
L.~WANG, L.~PANG and K.~ZHOU, \emph{{Applications of deep learning in high energy nuclear physics}}, \href{https://doi.org/https://doi.org/10.1360/SSPMA-2021-0300}{\emph{Sci. China Phys. Mech. Astron.} {\bfseries 52} (2022) 252003}.

\bibitem{WanBing:2022ssp}
W.~HE, J.~HE, R.~Wang and Y.~MA, \emph{{Machine learning applications in nuclear physics}}, \href{https://doi.org/https://doi.org/10.1360/SSPMA-2021-0309}{\emph{Sci. China Phys. Mech. Astron.} {\bfseries 52} (2022) 252004}.

\bibitem{YiLun:2022ssp}
Y.-L.~DU, D.~PABLOS and K.~Tywoniuk, \emph{{Applications of deep learning in jet quenching}}, \href{https://doi.org/https://doi.org/10.1360/SSPMA-2022-0046}{\emph{Sci. China Phys. Mech. Astron.} {\bfseries 52} (2022) 252017}.

\bibitem{fupeng:2022ssp}
F.~LI, L.~PANG and X.~WANG, \emph{{Application of machine learning to the study of QCD transition in heavy ion collisions}}, \href{https://doi.org/10.11889/j.0253-3219.2023.hjs.46.040014}{\emph{Nucl. Sci. Tech.} {\bfseries 46} (2023) 040014}.

\bibitem{He:2023urp}
W.~He, Q.~Li, Y.~Ma, Z.~Niu, J.~Pei and Y.~Zhang, \emph{{Machine learning in nuclear physics at low and intermediate energies}}, \href{https://doi.org/10.1007/s11433-023-2116-0}{\emph{Sci. China Phys. Mech. Astron.} {\bfseries 66} (2023) 282001} [\href{https://arxiv.org/abs/2301.06396}{{\ttfamily 2301.06396}}].

\bibitem{Ma:2023zfj}
Y.-G.~Ma, L.-G.~Pang, R.~Wang and K.~Zhou, \emph{{Phase Transition Study Meets Machine Learning}}, \href{https://doi.org/10.1088/0256-307X/40/12/122101}{\emph{Chin. Phys. Lett.} {\bfseries 40} (2023) 122101} [\href{https://arxiv.org/abs/2311.07274}{{\ttfamily 2311.07274}}].

\bibitem{Zhou:2023pti}
K.~Zhou, L.~Wang, L.-G.~Pang and S.~Shi, \emph{{Exploring QCD matter in extreme conditions with Machine Learning}}, \href{https://doi.org/10.1016/j.ppnp.2023.104084}{\emph{Prog. Part. Nucl. Phys.} {\bfseries 135} (2024) 104084} [\href{https://arxiv.org/abs/2303.15136}{{\ttfamily 2303.15136}}].

\bibitem{Boehnlein:2021eym}
A.~Boehnlein et~al., \emph{{Colloquium: Machine learning in nuclear physics}}, \href{https://doi.org/10.1103/RevModPhys.94.031003}{\emph{Rev. Mod. Phys.} {\bfseries 94} (2022) 031003} [\href{https://arxiv.org/abs/2112.02309}{{\ttfamily 2112.02309}}].

\bibitem{Aarts:2025gyp}
G.~Aarts, K.~Fukushima, T.~Hatsuda, A.~Ipp, S.~Shi, L.~Wang et~al., \emph{Physics-{Driven} {Learning} for {Inverse} {Problems} in {Quantum} {Chromodynamics}}, \href{https://doi.org/10.1038/s42254-024-00798-x}{\emph{Nature Reviews Physics} {\bfseries 7} (2025) 154}.

\bibitem{point-m2ae}
R.~Zhang, Z.~Guo, P.~Gao, R.~Fang, B.~Zhao, D.~Wang et~al., \emph{Point-m2ae: multi-scale masked autoencoders for hierarchical point cloud pre-training}, {\emph{Advances in neural information processing systems} {\bfseries 35} (2022) 27061}.

\bibitem{2d23d}
R.~Zhang, L.~Wang, Y.~Qiao, P.~Gao and H.~Li, \emph{Learning 3d representations from 2d pre-trained models via image-to-point masked autoencoders},  in \emph{Proceedings of the IEEE/CVF Conference on Computer Vision and Pattern Recognition}, pp.~21769--21780, 2023.

\bibitem{point-bert}
X.~Yu, L.~Tang, Y.~Rao, T.~Huang, J.~Zhou and J.~Lu, \emph{Point-bert: Pre-training 3d point cloud transformers with masked point modeling},  in \emph{Proceedings of the IEEE/CVF conference on computer vision and pattern recognition}, pp.~19313--19322, 2022.

\bibitem{maskpoint}
H.~Liu, M.~Cai and Y.J.~Lee, \emph{Masked discrimination for self-supervised learning on point clouds},  in \emph{European Conference on Computer Vision}, pp.~657--675, Springer, 2022.

\bibitem{golling2024masked}
T.~Golling, L.~Heinrich, M.~Kagan, S.~Klein, M.~Leigh, M.~Osadchy et~al., \emph{Masked particle modeling on sets: towards self-supervised high energy physics foundation models}, {\emph{Machine Learning: Science and Technology} {\bfseries 5} (2024) 035074}.

\bibitem{birk2024omnijet}
J.~Birk, A.~Hallin and G.~Kasieczka, \emph{Omnijet-$\alpha$: the first cross-task foundation model for particle physics}, {\emph{Machine Learning: Science and Technology} {\bfseries 5} (2024) 035031}.

\bibitem{harris2024re}
P.~Harris, M.~Kagan, J.~Krupa, B.~Maier and N.~Woodward, \emph{Re-simulation-based self-supervised learning for pre-training foundation models}, {\emph{arXiv preprint arXiv:2403.07066} (2024) }.

\bibitem{mikuni2025method}
V.~Mikuni and B.~Nachman, \emph{Method to simultaneously facilitate all jet physics tasks}, {\emph{Physical Review D} {\bfseries 111} (2025) 054015}.

\bibitem{lin2005multiphase}
Z.-W.~Lin, C.M.~Ko, B.-A.~Li, B.~Zhang and S.~Pal, \emph{Multiphase transport model for relativistic heavy ion collisions}, {\emph{Physical Review C—Nuclear Physics} {\bfseries 72} (2005) 064901}.

\bibitem{lin2021further}
Z.-W.~Lin and L.~Zheng, \emph{Further developments of a multi-phase transport model for relativistic nuclear collisions}, {\emph{Nuclear Science and Techniques} {\bfseries 32} (2021) 113}.

\bibitem{pointnet}
C.R.~Qi, H.~Su, K.~Mo and L.J.~Guibas, \emph{Pointnet: Deep learning on point sets for 3d classification and segmentation},  in \emph{Proceedings of the IEEE conference on computer vision and pattern recognition}, pp.~652--660, 2017.

\bibitem{guo2024machine}
S.~Guo, H.-S.~Wang, K.~Zhou and G.-L.~Ma, \emph{Machine learning study to identify collective flow in small and large colliding systems}, {\emph{Physical Review C} {\bfseries 110} (2024) 024910}.

\bibitem{lai2022information}
Y.S.~Lai, J.~Mulligan, M.~P{\l}osko{\'n} and F.~Ringer, \emph{The information content of jet quenching and machine learning assisted observable design}, {\emph{Journal of High Energy Physics} {\bfseries 2022} (2022) 1}.

\bibitem{CMS:2013jlh}
{\scshape CMS} collaboration, \emph{{Multiplicity and Transverse Momentum Dependence of Two- and Four-Particle Correlations in pPb and PbPb Collisions}}, \href{https://doi.org/10.1016/j.physletb.2013.06.028}{\emph{Phys. Lett. B} {\bfseries 724} (2013) 213} [\href{https://arxiv.org/abs/1305.0609}{{\ttfamily 1305.0609}}].

\bibitem{NIPS2017_8a20a862}
S.M.~Lundberg and S.-I.~Lee, \emph{A unified approach to interpreting model predictions},  in \emph{Advances in Neural Information Processing Systems}, I.~Guyon, U.V.~Luxburg, S.~Bengio, H.~Wallach, R.~Fergus, S.~Vishwanathan et~al., eds., vol.~30, Curran Associates, Inc., 2017.

\bibitem{dohi2020variational}
K.~Dohi, \emph{Variational autoencoders for jet simulation}, {\emph{arXiv preprint arXiv:2009.04842} (2020) }.

\bibitem{ostdiek2022deep}
B.~Ostdiek, \emph{Deep set auto encoders for anomaly detection in particle physics}, {\emph{SciPost Physics} {\bfseries 12} (2022) 045}.

\bibitem{mikuni2023fast}
V.~Mikuni, B.~Nachman and M.~Pettee, \emph{Fast point cloud generation with diffusion models in high energy physics}, {\emph{Physical Review D} {\bfseries 108} (2023) 036025}.

\bibitem{butter2025jet}
A.~Butter, N.~Huetsch, S.~Palacios~Schweitzer, T.~Plehn, P.~Sorrenson and J.~Spinner, \emph{Jet diffusion versus jetgpt--modern networks for the lhc}, {\emph{SciPost Physics Core} {\bfseries 8} (2025) 026}.

\bibitem{qu2022particle}
H.~Qu, C.~Li and S.~Qian, \emph{Particle transformer for jet tagging},  in \emph{International Conference on Machine Learning}, pp.~18281--18292, PMLR, 2022.

\bibitem{qiu2023holistic}
S.~Qiu, S.~Han, X.~Ju, B.~Nachman and H.~Wang, \emph{Holistic approach to predicting top quark kinematic properties with the covariant particle transformer}, {\emph{Physical Review D} {\bfseries 107} (2023) 114029}.

\bibitem{bogatskiy2022symmetry}
A.~Bogatskiy, S.~Ganguly, T.~Kipf, R.~Kondor, D.W.~Miller, D.~Murnane et~al., \emph{Symmetry group equivariant architectures for physics}, {\emph{arXiv preprint arXiv:2203.06153} (2022) }.

\bibitem{gambhir2024streamlining}
R.~Gambhir, A.~Osathapan and J.~Thaler, \emph{Streamlining latent spaces in machine learning using moment pooling}, {\emph{Physical Review D} {\bfseries 110} (2024) 074020}.

\bibitem{hallin2025universal}
A.~Hallin, G.~Kasieczka, S.~Kraml, A.~Lessa, L.~Moureaux, T.~von Schwartz et~al., \emph{Universal new physics latent space}, {\emph{Physical Review D} {\bfseries 111} (2025) 016006}.

\bibitem{gong2022efficient}
S.~Gong, Q.~Meng, J.~Zhang, H.~Qu, C.~Li, S.~Qian et~al., \emph{An efficient lorentz equivariant graph neural network for jet tagging}, {\emph{Journal of High Energy Physics} {\bfseries 2022} (2022) 1}.

\bibitem{bogatskiy2022pelican}
A.~Bogatskiy, T.~Hoffman, D.W.~Miller and J.T.~Offermann, \emph{Pelican: Permutation equivariant and lorentz invariant or covariant aggregator network for particle physics}, {\emph{arXiv preprint arXiv:2211.00454} (2022) }.

\bibitem{spinner2024lorentz}
J.~Spinner, V.~Bres{\'o}, P.~De~Haan, T.~Plehn, J.~Thaler and J.~Brehmer, \emph{Lorentz-equivariant geometric algebra transformers for high-energy physics}, {\emph{Advances in neural information processing systems} {\bfseries 37} (2024) 22178}.

\bibitem{brehmer2024lorentz}
J.~Brehmer, V.~Bres{\'o}, P.~de~Haan, T.~Plehn, H.~Qu, J.~Spinner et~al., \emph{A lorentz-equivariant transformer for all of the lhc}, {\emph{arXiv preprint arXiv:2411.00446} (2024) }.

\bibitem{spinner2025lorentz}
J.~Spinner, L.~Favaro, P.~Lippmann, S.~Pitz, G.~Gerhartz, T.~Plehn et~al., \emph{Lorentz local canonicalization: How to make any network lorentz-equivariant}, {\emph{arXiv preprint arXiv:2505.20280} (2025) }.

\bibitem{favaro2025lorentz}
L.~Favaro, G.~Gerhartz, F.A.~Hamprecht, P.~Lippmann, S.~Pitz, T.~Plehn et~al., \emph{Lorentz-equivariance without limitations}, {\emph{arXiv preprint arXiv:2508.14898} (2025) }.

\bibitem{Fukushima:2008xe}
K.~Fukushima, D.E.~Kharzeev and H.J.~Warringa, \emph{{The Chiral Magnetic Effect}}, \href{https://doi.org/10.1103/PhysRevD.78.074033}{\emph{Phys. Rev. D} {\bfseries 78} (2008) 074033} [\href{https://arxiv.org/abs/0808.3382}{{\ttfamily 0808.3382}}].

\bibitem{Zhao:2021yjo}
Y.-S.~Zhao, L.~Wang, K.~Zhou and X.-G.~Huang, \emph{{Detecting the chiral magnetic effect via deep learning}}, \href{https://doi.org/10.1103/PhysRevC.106.L051901}{\emph{Phys. Rev. C} {\bfseries 106} (2022) L051901} [\href{https://arxiv.org/abs/2105.13761}{{\ttfamily 2105.13761}}].

\bibitem{Guo:2025wry}
S.~Guo, L.~Wang, K.~Zhou and G.-L.~Ma, \emph{{Neural Unfolding of the Chiral Magnetic Effect in Heavy-Ion Collisions}},  \href{https://arxiv.org/abs/2507.05808}{{\ttfamily 2507.05808}}.

\bibitem{Jia:2022qgl}
J.~Jia, G.~Giacalone and C.~Zhang, \emph{{Separating the Impact of Nuclear Skin and Nuclear Deformation in High-Energy Isobar Collisions}}, \href{https://doi.org/10.1103/PhysRevLett.131.022301}{\emph{Phys. Rev. Lett.} {\bfseries 131} (2023) 022301} [\href{https://arxiv.org/abs/2206.10449}{{\ttfamily 2206.10449}}].

\bibitem{Zhao:2022grq}
X.-L.~Zhao and G.-L.~Ma, \emph{{Search for the chiral magnetic effect in collisions between two isobars with deformed and neutron-rich nuclear structures}}, \href{https://doi.org/10.1103/PhysRevC.106.034909}{\emph{Phys. Rev. C} {\bfseries 106} (2022) 034909} [\href{https://arxiv.org/abs/2203.15214}{{\ttfamily 2203.15214}}].

\bibitem{MA4pictures}
K.~He, X.~Chen, S.~Xie, Y.~Li, P.~Dollar and R.~Girshick, \emph{Masked autoencoders are scalable vision learners},  in \emph{Proceedings - 2022 IEEE/CVF Conference on Computer Vision and Pattern Recognition, CVPR 2022}, Proceedings of the IEEE Computer Society Conference on Computer Vision and Pattern Recognition, pp.~15979--15988, IEEE Computer Society, 2022, \href{https://doi.org/10.1109/CVPR52688.2022.01553}{DOI}.

\end{thebibliography}\endgroup
\appendix

\section{Technical Appendices and Supplementary Material}

\begin{table}[ht]
	\caption{Pretraining stage training hyperparameter settings}
	\label{tab1}
	\centering % 保持居中
	\begin{tabular}{ll} % 修改为l（左对齐）或c（居中）
		\toprule
		hyperparameter & value \\
		\midrule
		epochs   & 300                         \\
		optimizer  & AdamW                         \\
		learning rate & 0.001                     \\
		weight decay   & 0.05                          \\
		LR schedule    & cosine decay \\
        warmup epochs   & 3 \\
        batch size    & 256 \\
        mask ratio    & 0.25 \\
        mask type    & random \\
		\bottomrule
	\end{tabular}
\end{table}

\begin{table}[ht]
	\caption{Finetuning phase training hyperparameter settings}
	\label{tab2}
	\centering % 保持居中
	\begin{tabular}{ll} % 修改为l（左对齐）或c（居中）
		\toprule
		hyperparameter & value \\
		\midrule
		epochs    & 300                         \\
		optimizer  & AdamW                         \\
		learning rate & 0.0005                     \\
		weight decay   & 0.05                          \\
		LR schedule    & cosine decay \\
        warmup epochs    & 10 \\
        batch size    & 128 \\
		\bottomrule
	\end{tabular}
\end{table}

\begin{table}[h]
\centering
\caption{{Ablation studies: (1) Effect of PointNet preprocessing; (2) Effect of different masking ratios.}}
\label{tab:ablation}
\begin{tabular}{lcc}
\hline
\textbf{Experiment} & \textbf{Setting} & \textbf{Accuracy} \\
\hline
PointNet Preprocessing & With PointNet  & 0.9775 \\
                       & Without PointNet & 0.7303 \\
\hline
Masking Ratio          & 25\% & 0.9775 \\
                       & 50\% & 0.7853 \\
                       & 75\% & 0.7232 \\
\hline
\end{tabular}
\end{table}

\begin{table}[ht]
\centering
\caption{{Number of parameters for different models. Note that while the pretraining stage includes a larger model due to the decoder, the finetuned model  has a parameter count comparable to the PointNet baseline. The observed performance gains thus primarily arise from the two-stage pretraining and finetuning strategy rather than model scale.}}
\begin{tabular}{lccc}
\toprule
Model & Stage & Total Parameters & Trainable Parameters \\
\midrule
encoder + decoder & Pretrain & 2.32M & 1.16M \\
encoder + downstream head & Finetune & 1.07M & 1.07M \\
PointNet & Baseline & 0.74M & 0.74M \\
\bottomrule
\end{tabular}
\label{tab:model_parameters}
\end{table}

\subsection*{Hardware and Training Time}
All experiments were conducted on a laptop equipped with an NVIDIA RTX 4070 GPU (8 GB VRAM), an Intel i7-12800HX CPU, and 32 GB of system memory. 
Training for 300 epochs required approximately 45,600 seconds.

\subsection*{Code Availability}
The implementation of our model, including pre-training and fine-tuning procedures, is publicly available at: 
\url{https://github.com/Giovanni-Sforza/MaskPoint-AMPT}.

\subsection{{Justification of Pre-Training Design Choices}}
{Here, we elaborate on the key design choices made during the self-supervised pre-training phase.}

\subsubsection{Discrimination vs. Reconstruction Task}
{In the domain of image processing, masked autoencoders often follow a reconstruction paradigm: they apply patch-wise masking to an input image, retain the positional encodings of the masked patches, and then use a decoder to reconstruct the RGB values of the masked pixels from the visible regions ~\citep{MA4pictures}. The underlying assumption is that the geometric position of a pixel is known framework information, while its color is the core information to be modeled.}

{However, this paradigm is ill-suited for relativistic heavy-ion collisions. In this context, the momentum-space coordinates of the final-state particles constitute the core physical information, as their distribution is directly linked to the initial thermodynamic properties of the collision. A direct application of the image reconstruction paradigm would allow the decoder to simply copy the positional encodings to reconstruct the momentum coordinates, causing the model to degenerate into a trivial memorization task without learning deep physical features.}

{To overcome this challenge, we adopted a discrimination-based pre-training scheme ~\citep{ maskpoint}. As illustrated in Figure \ref{fig.autoencoder}, this approach tasks the model with distinguishing between the "real" masked particles and randomly sampled "fake" particles. This forces the encoder to learn the underlying distribution and correlations within the particle point cloud to successfully identify what constitutes a physically plausible particle within the context of the event, thereby learning a much richer and more meaningful feature representation.}

\subsubsection{Sampling Strategy: Farthest Point Sampling} 
{We chose Farthest Point Sampling (FPS) over a uniform random sampling strategy to select the 25\% of points to be masked. FPS iteratively selects points that are maximally distant from the set of points already chosen. This ensures that the unmasked points provide a more uniform coverage of the entire point cloud's geometric structure.}

\subsubsection{Dataset and Training Procedure}
{Both the self-supervised pre-training and the supervised fine-tuning stages utilized the entire available AMPT-generated dataset. The models were trained for a maximum of 300 epochs. We employed a checkpointing strategy, saving the model weights after each epoch if the performance on a validation set improved. The final models used for analysis correspond to the epoch with the best validation performance.}

\end{document}